\documentclass[pra,amssymb,amsmath]{revtex4}

\usepackage{appendix}
\usepackage{graphicx}
\usepackage{amssymb,amsmath,amsbsy} 

\usepackage[polish,english]{babel}
\usepackage[T1]{fontenc}

\def\B#1{\!\left(#1\right)}
\def\BB#1{\!\left[#1\right]}

\def\be{\begin{equation}}
\def\ee{\end{equation}}

\def\bee{\begin{equation*}}
\def\eee{\end{equation*}}

\def\bx{\begin{equation}\begin{aligned}}
\def\ex{\end{aligned}\end{equation}}

\def\bxx{\begin{equation*}\begin{aligned}}
\def\exx{\end{aligned}\end{equation*}}

\def\bg{\begin{equation}\begin{gathered}}
\def\eg{\end{gathered}\end{equation}}

\def\bgg{\begin{equation*}\begin{gathered}}
\def\egg{\end{gathered}\end{equation*}}

\def\kA{ \dfrac{2j+1}{2n}\pi }
\def\kB{ \dfrac{j\pi}{n} }
\def\kC{ \dfrac{2j+1}{2n+1}\pi }
\def\kD{ \dfrac{2j\pi}{2n+1} }

\def\xdwa{\dfrac{x}{2}}
\def\pol{\dfrac{1}{2}}

\def\la{\langle}
\def\ra{\rangle}

\begin{document}

\title{The quantum Ising model: finite sums and hyperbolic functions}

\author{Bogdan Damski}
\affiliation{Jagiellonian University, Institute of Physics, {\L}ojasiewicza 11, 30-348 Krak\'ow, Poland}

\begin{abstract}
We derive exact  closed-form expressions for several  sums leading to hyperbolic
functions and discuss their applicability for studies of finite-size Ising spin chains.
We show how they immediately   lead to 
closed-form expressions for both fidelity susceptibility characterizing the quantum critical
point and the coefficients of the counterdiabatic Hamiltonian  enabling arbitrarily quick
adiabatic driving of the system.
Our results generalize and extend the sums presented in the
popular  Gradshteyn and Ryzhik Table of Integrals, Series, and Products. 
\end{abstract}
\maketitle

\section*{Introduction}

In typical condensed matter setups the size of a system
is thermodynamic and so the finite-size effects are negligible.  
The situation can be quite the opposite with quantum simulators, i.e., 
quantum systems such as cold atoms \cite{MaciekBook,GreinerNat2011} and ions \cite{Qsim1,Qsim2}
that are tailored to emulate the challenging condensed matter models. 
One of the ultimate goals of this field is to emulate spin systems that are
analytically and numerically unapproachable in the thermodynamic limit.
Realization of such a goal  requires substantial experimental progress, 
but it is reasonable to expect that on-demand emulation of arbitrary spin models
is within experimental reach \cite{Monroe0}.
The current state-of-the-art
experiments, however,  manage to emulate  spin systems 
composed of less than $20$  spins \cite{LanyonSci2011,Monroe1,Monroe2}, where the finite-size effects come into play.

It is thus desirable to develop better understanding of finite-size effects 
in many-body spin systems to be able to (i) extrapolate the finite-size data
to the thermodynamic limit; (ii)  understand  the limitations of the current
generation of quantum emulators.
Actually, the finite-size effects are interesting on their own in systems undergoing a 
quantum phase transition. Indeed, such systems are 
especially sensitive to the system size because their correlation length diverges 
around the critical point in the thermodynamic limit \cite{Sachdev,SachdevToday}. 
Additionally, we mention that
finite-size systems are relevant for quantum engineering, where the goal is to 
prepare  a system in a desired quantum state (see e.g. Ref.
\cite{AdolfoAdv2013}).

Due to its exact solvability, the one dimensional quantum Ising model 
is a perfect system for analytical studies of a quantum phase transition \cite{Sachdev,SachdevToday}. 
Such studies have been extensively performed in the past in the
thermodynamic limit. This limit  allowed for the observation of 
singularities at the critical point and simplified the analytical
computations \cite{Lieb1961,BarouchPRA1971,Pfeuty}. 

The calculations in the thermodynamic limit of the quantum
Ising model rely on the  replacement of sums with integrals. 
If we would be able to find closed-form expressions for such  sums, we would
get detailed analytical insights into the finite-size effects in a model system 
undergoing a quantum phase transition.
The main goal of this work is to provide several  closed-form expressions for the sums 
that appear in the studies of the quantum Ising model. Additionally, 
we briefly discuss two
examples, where the use of our sums immediately leads to  elegant 
expressions for both  
fidelity susceptibility capturing the key properties of the quantum
critical point and the coefficients of the counterdiabatic  Hamiltonian allowing for
implementation of  perfectly adiabatic dynamics  on an 
arbitrarily short time scale.

\section*{Results}
\label{sec_Sums}
There are four interestingly looking sums in Chapter 1.382 of Ref. \cite{Ryzhik}:
\bg
\label{R1}
\sum_{j=0}^{n-1}\dfrac{\sinh(x)}{\sin^2\B{\dfrac{2j+1}{4n}\pi}+\sinh^2\B{\xdwa}}=2n\tanh(nx),
\eg
\bg
\label{R2}
\sum_{j=1}^{n-1}\dfrac{\sinh(x)}{\sin^2\B{\dfrac{j\pi}{2n}}+\sinh^2\B{\xdwa}}=2n\coth(nx)-2\coth(x),
\eg
\bg
\label{R3}
\sum_{j=0}^{n-1}\dfrac{\sinh(x)}{\sin^2\B{\dfrac{2j+1}{2(2n+1)}\pi}+\sinh^2\B{\xdwa}}=
(2n+1)\tanh\BB{\B{n+\pol}x}-\tanh\B{\xdwa},
\eg
\bg
\label{R4}
\sum_{j=1}^{n}\dfrac{\sinh(x)}{\sin^2\B{\dfrac{j\pi}{2n+1}}+\sinh^2\B{\xdwa}}=
(2n+1)\coth\BB{\B{n+\pol}x}-\coth\B{\dfrac{x}{2}},
\eg
which are valid for integer $n\ge1$ (we assume such $n$ everywhere in this
article).
To explain where these sums come from, we consider  polynomials
\bgg
g^{2n}+1=\prod_{j=-n}^{n-1}\BB{g-\exp\B{i\kA}}=\prod_{j=0}^{n-1}\BB{g^2-2g\cos\B{\kA}+1},
\egg
\bgg
g^{2n}-1=\prod_{j=-n}^{n-1}\BB{g-\exp\B{i\dfrac{j\pi}{n}}}=(g^2-1)\prod_{j=1}^{n-1}
\BB{g^2-2g\cos\B{\dfrac{j\pi}{n}}+1},
\egg
\bgg
g^{2n+1}+1=\prod_{j=-n}^n\BB{g-\exp\B{i\kC}}=(g+1)
\prod_{j=0}^{n-1}\BB{g^2-2g\cos\B{\kC}+1},
\egg
\bgg
g^{2n+1}-1=\prod_{j=-n}^n\BB{g-\exp\B{i\dfrac{2j\pi}{2n+1}}}=
(g-1)\prod_{j=1}^{n}\BB{g^2-2g\cos\B{\dfrac{2j\pi}{2n+1}}+1},
\egg
which can be also written as 
\bg
\label{t1}
\ln\B{g^{2n}+1}=\sum_{j=0}^{n-1}\ln\BB{g^2-2g\cos\B{\kA}+1},
\eg
\bg
\label{t2}
\ln\dfrac{g^{2n}-1}{g^2-1}=\sum_{j=1}^{n-1}\ln\BB{g^2-2g\cos\B{\dfrac{j\pi}{n}}+1},
\eg
\bg
\label{t3}
\ln\dfrac{g^{2n+1}+1}{g+1}=\sum_{j=0}^{n-1}\ln\BB{g^2-2g\cos\B{\kC}+1},
\eg
\bg
\label{t4}
\ln\dfrac{g^{2n+1}-1}{g-1}=\sum_{j=1}^{n}\ln\BB{g^2-2g\cos\B{\dfrac{2j\pi}{2n+1}}+1}.
\eg
If we now substitute 
\be
g=\exp(x)
\label{g}
\ee
into Eqs. (\ref{t1})--(\ref{t4}) and 
differentiate them   with respect to $x$, we will get Eqs.
(\ref{R1})--(\ref{R4}) after simple algebra. As will be explained in the
Discussion section, the parameter $g$ (\ref{g}) is the magnetic field 
in the quantum Ising model.

The sums (\ref{R1})--(\ref{R4}) can be generalized in the following way 
\bg
\sum_{j=0}^{n-1}\dfrac{\sinh(x)\cos\B{m\kA}}{\sin^2\B{\dfrac{2j+1}{4n}\pi}+\sinh^2\B{\xdwa}}
= 2n\dfrac{\sinh\BB{(n-m)x}}{\cosh(nx)}, 
\label{RB1}
\eg
\bg
\sum_{j=1}^{n-1}\dfrac{\sinh(x)\cos\B{m\kB}}{\sin^2\B{\dfrac{j\pi}{2n}}+\sinh^2\B{\xdwa}}=
2n\dfrac{\cosh\BB{(n-m)x}}{\sinh(nx)}-2\dfrac{\cosh\BB{\dfrac{1+(-1)^m}{2}x}}{\sinh(x)},
\label{RB2}
\eg
\bg
\sum_{j=0}^{n-1}\dfrac{\sinh(x)\cos\B{m\kC}}{\sin^2\B{\dfrac{2j+1}{2(2n+1)}\pi}+\sinh^2\B{\xdwa}}=
(2n+1)\dfrac{\sinh\BB{\B{n-m+\pol}x}}{\cosh\BB{\B{n+\pol}x}}-(-1)^m\tanh\B{\dfrac{x}{2}},
\label{RB3}
\eg
\bg
\sum_{j=1}^{n}\dfrac{\sinh(x)\cos\B{m\kD}}{\sin^2\B{\dfrac{j\pi}{2n+1}}+\sinh^2\B{\xdwa}}=
\B{2n+1}\dfrac{\cosh\BB{\B{n-m+\pol}x}}{\sinh\BB{\B{n+\pol}x}}-\coth\B{\xdwa}.
\label{RB4}
\eg
Eqs. (\ref{RB1})--(\ref{RB2}) are valid for $m=0,1,\cdots,2n$, while 
Eqs. (\ref{RB3})--(\ref{RB4}) hold for $m=0,1,\cdots,2n+1$. 
These results reproduce Eqs. (\ref{R1})--(\ref{R4}) for $m=0$.

Additionally, we propose another set of identities 
\bg
\sum_{j=0}^{n-1}\dfrac{\sin\B{\kA}\sin\B{m\kA}}{\sin^2\B{\dfrac{2j+1}{4n}\pi}+\sinh^2\B{\xdwa}}
= 2n\dfrac{\cosh\BB{(n-m)x}}{\cosh(nx)},
\label{B1}
\eg
\bg
\sum_{j=1}^{n-1}\dfrac{\sin\B{\kB}\sin\B{m\kB}}{\sin^2\B{\dfrac{j\pi}{2n}}+\sinh^2\B{\xdwa}}=
2n\dfrac{\sinh\BB{(n-m)x}}{\sinh(nx)}, 
\label{B2}
\eg
\bg
\sum_{j=0}^{n-1}\dfrac{\sin\B{\kC}\sin\B{m\kC}}{\sin^2\B{\dfrac{2j+1}{2(2n+1)}\pi}+\sinh^2\B{\xdwa}}=
(2n+1)\dfrac{\cosh\BB{\B{n-m+\pol}x}}{\cosh\BB{\B{n+\pol}x}},
\label{B3}
\eg
\bg
\sum_{j=1}^{n}\dfrac{\sin\B{\kD}\sin\B{m\kD}}{\sin^2\B{\dfrac{j\pi}{2n+1}}+\sinh^2\B{\xdwa}}=
(2n+1)\dfrac{\sinh\BB{\B{n-m+\pol}x}}{\sinh\BB{\B{n+\pol}x}}.
\label{B4}
\eg
Eqs. (\ref{B1}) and (\ref{B2}) yield a correct result for $m=1,2,\cdots,2n-1$,
while Eqs. (\ref{B3})--(\ref{B4}) hold for $m=1,2,\cdots,2n$.

The sums (\ref{RB1})--(\ref{B4}) have been
conjectured by applying the techniques described in Appendix A of Ref.
\cite{BD_counter}. Two of them, (\ref{RB1}) and (\ref{B1}), have been already
proven in Ref. \cite{BD_counter}. We will outline below the steps allowing for
an inductive proof of all the sums. This way we will  generalize the former results and 
simplify a bit the proof. 

To compactify notation, we define two functions
\be
f_m(x)=\sum_k\dfrac{\sinh(x)\cos\B{mk}}{\sin^2\B{k/2}+\sinh^2\B{x/2}},
\label{fm}
\ee
\be
h_m(x)=
\sum_k\dfrac{\sin\B{k}\sin\B{mk}}{\sin^2\B{k/2}+\sinh^2\B{x/2}},
\label{hm}
\ee
where the momenta $k$ are chosen such that $\sum_k$ replaces $\sum_j$ in 
Eqs. (\ref{RB1})--(\ref{B4}). Next, we show that
\bx
f_{m+1}(x)=&\cosh(x)f_m(x)-\sinh(x)h_m(x)
           -\sinh(x)\Delta_m,\\
h_{m+1}(x)=&\cosh(x)h_m(x)-\sinh(x)f_m(x)+\cosh(x)\Delta_m+\Delta_{m+1},
\label{rec}
\ex
where $\Delta_m=2\sum_k\cos(mk)$. These recurrence relations 
can be derived by substituting 
\be
\cos(k)=\cosh(x)-2[\sin^2(k/2)+\sinh^2(x/2)]
\label{cosk}
\ee
and 
\bx
\sin^2(k)=&2[\cosh(x)+\cos(k)][\sin^2(k/2)+\sinh^2(x/2)]
-\sinh^2(x)
\label{sin2k}
\ex
into $\cos\BB{(m+1)k}=\cos(mk)\cos(k)-\sin(mk)\sin(k)$ and 
$\sin(k)\sin\BB{(m+1)k}=\sin(k)\sin(mk)\cos(k)+\cos(mk)\sin^2(k)$.

To proceed with the inductive proof, we find in an elementary way that  
\bx
\Delta_m&=0,  \  k=\kA,   \ j= 0,1,\cdots,n-1,\\
\Delta_m&=(-1)^{m+1}-1, \  k=\kB, \   j=1,2,\cdots,n-1,
\label{Dm1}
\ex
for $m=1,2,\cdots,2n-1$ and
\bx
\label{Dm2}
\Delta_m&=(-1)^{m+1}, \ k=\kC,  \ j=0,1,\cdots,n-1,\\
\Delta_m&=-1,  \ k=\kD, \ j=1,2,\cdots,n,
\ex
for $m=1,2,\cdots,2n$.

First, we assume that $f_m(x)$
and $h_m(x)$ are given by the right-hand sides of Eqs.
(\ref{RB1})--(\ref{RB4}) and (\ref{B1})--(\ref{B4}), respectively.

Second, we check that our inductive conjecture holds for  $m=1$. This is done
by substituting Eqs. (\ref{cosk}) and (\ref{sin2k}) into 
Eqs. (\ref{fm}) and (\ref{hm}) taken at $m=1$ and then comparing the 
resulting expressions to Eqs. (\ref{R1})--(\ref{R4}).

Third, it is a straightforward exercise  to check that the expressions for
$f_{m+1}(x)$ and $h_{m+1}(x)$ obtained from the recurrence equations (\ref{rec}) 
are exactly the same as the right-hand sides of the conjectured equations  
(\ref{RB1})--(\ref{B4}) after the replacement of $m$ by $m+1$.
The last remark is true provided that expressions for $\Delta_m$ 
are used in the recurrence equations for appropriate $m$ listed below Eqs.
(\ref{Dm1}) and (\ref{Dm2}). 
This completes the  proof that 
Eqs. (\ref{RB1}), (\ref{RB2}), (\ref{B1}),  and (\ref{B2}) are correct for $m=1,2,\cdots,2n-1$, while 
Eqs. (\ref{RB3}), (\ref{RB4}), (\ref{B3}), and (\ref{B4})  hold for 
 $m=1,2,\cdots,2n$. 
Additionally, one may substitute $m=2n$ into Eqs. (\ref{RB1}) 
and (\ref{RB2}) and $m=2n+1$ into Eqs. (\ref{RB3}) and (\ref{RB4}) to 
immediately verify with Eqs. (\ref{R1})--(\ref{R4}) that Eqs.
(\ref{RB1})--(\ref{RB4}) are correct for such $m$  as well. This concludes
the  proof of Eqs. (\ref{RB1})--(\ref{B4}).

\section*{Discussion}
The results obtained in the previous section provide useful analytical tools
for the studies of the quantum Ising model. The Hamiltonian of this model reads
\be
\hat H = -\sum_{i=1}^N \B{\sigma^x_i\sigma^x_{i+1} + g\sigma^z_i},
\label{HIsing}
\ee
where $N$ is the number of spins, $\vec{\sigma}_i$ denote the  Pauli matrices 
acting on the $i$-th spin,  $g$ is a magnetic field, and periodic boundary conditions 
are assumed $\vec{\sigma}_{i+N}=\vec{\sigma_i}$. 
This exactly solvable model has many interesting properties \cite{Sachdev,SachdevToday}. We will
mention two of them that are relevant for the present discussion.

First, the above Hamiltonian commutes with the so-called parity operator
\bee
\hat P =  \prod_{i=1}^N \sigma^z_i.
\eee
Therefore, its Hilbert space can be divided into positive and negative parity subspaces (eigenvalues of
the parity operator are $\pm1$ because $\hat P^2=1$).
The parity of the ground state depends on both the system size and the
direction of the magnetic field (see e.g. Ref. \cite{BD_fidJPA}). 

Second, the quantum Ising model can be mapped to the non-interacting lattice  fermionic  model 
through the Jordan-Wigner transformation and then it can be diagonalized 
in the momentum space \cite{Lieb1961,BarouchPRA1970,Sachdev}.
There are four possibilities for the choice of momenta $k$
in a periodic system (see e.g. Ref. \cite{BD_fidJPA}). In the positive 
parity subspace of the even-sized system
\be
k=\pm\kA, \  \  j=0,1,\cdots,n-1,  \ \ N=2n,
\label{KA}
\ee
while in the negative parity subspace of such a system
\be
k=0,\pm\kB,\pi, \ \  j=1,2,\cdots,n-1, \ \  N=2n.
\label{KB}
\ee
In the positive parity subspace of the odd-sized system 
\be
k=\pm\kC,\pi,  \ \ j=0,1,\cdots,n-1, \ \ N=2n+1,
\label{KC}
\ee
while in the negative parity subspace of such a system 
\be
k=0,\pm\kD,\  \  j=1,2,\cdots,n, \ \  N=2n+1.
\label{KD}
\ee
Forgetting about the irrelevant sign of $k$ and the $0$ and $\pi$  momentum modes, 
the summations over these  four choices of momenta correspond 
to the  four summations over $j$ seen in Eqs. (\ref{RB1})--(\ref{RB4}) and
(\ref{B1})--(\ref{B4}).

What's more, the denominator of all the sums that we have computed, 
$\sin^2(k/2)+\sinh^2(x/2)$,
can be mapped to the form familiar from the Ising model computations,  
\be
g^2-2g\cos(k)+1, 
\label{gfact}
\ee
after setting 
\be
x=\ln(g).
\label{x}
\ee 
Note that we have already used this mapping at the beginning of the Results section (\ref{g}).
Since  both the eigenvalues and eigenstates of quantum Ising
model depend on the factor (\ref{gfact})-- see e.g. Ref. \cite{Sachdev} --
it is not surprising that the sums involving it appear in the studies of this model.

Finally, we note that the sums involving positive integer  powers of the factor (\ref{gfact}) 
in the denominator can be easily generated through the 
differentiation of the sums (\ref{RB1})--(\ref{B4})  with respect to $x$.
We are now ready  to show two applications for the results of the
previous section.

{\bf Fidelity approach to quantum phase transitions.} 
The location of the critical point and the critical exponent 
quantifying power-law divergence of the correlation length around the critical
point  can be extracted from the 
overlap between ground states corresponding to different external fields 
\cite{Zanardi,ABQ2010,Polkovnikov,Zhou,BD_fidPRL,GuReview}. 
Such an overlap is called fidelity \cite{Zanardi}. It defines the so-called fidelity 
susceptibility $\chi$ through a  Taylor expansion in the field shift $\delta$ \cite{You2007}
\be
\left|\la g|g+\delta\ra\right|=
1-\chi(g)\frac{\delta^2}{2}+ {\cal O}\B{\delta^3},
\label{fidelity}
\ee
where $|g\ra$ is a ground state of a Hamiltonian depending on an external
field $g$. It can be easily shown that \cite{Zanardi}
\be
\chi(g)=\frac{1}{4}\sum_{0<k<\pi} \frac{\sin^2(k)}{\B{g^2-2g\cos(k)+1}^2}
\label{chi}
\ee
in the quantum Ising model (\ref{HIsing}); the restriction to $0<k<\pi$ in the sum is applied
to momenta  from Eqs. (\ref{KA})--(\ref{KD}).

Therefore, to compute the exact closed-form expression for fidelity 
susceptibility $\chi(g)$ one needs to (i) determine the parity of the ground state and 
choose the appropriate momenta $k$  from the list (\ref{KA})--(\ref{KD});
(ii) choose the relevant sum from Eqs. (\ref{B1})--(\ref{B4}), set
$m=1$ to obtain $\sin^2(k)$ in the numerator of the summand, and compute
derivative of the resulting expression with respect to $x$ to square the summand's denominator;
(iii) substitute (\ref{x}) into the resulting expression. One gets in the end  
\be
\chi(g)= 
\frac{N^2}{16g^2}\frac{|g|^N}{\B{|g|^N+1}^2}+\frac{N}{16g^2}
\frac{|g|^N-g^2}{\B{|g|^N+1}\B{g^2-1}}
\label{chii}
\ee
for all system sizes and magnetic fields. This  expression was computed 
in a bit more complicated way in Refs. \cite{BD_fidJPA,BD_fidPRE}.

The quantum Ising model has the critical point at $g_c=1$ separating the ferromagnetic 
phase ($0\le g<1$) from the paramagnetic phase ($g>1$). It is thus expected
that its ground states are most sensitive to the change of the magnetic field
near the critical point. Therefore,  fidelity
(\ref{fidelity}) should have minimum near the critical point, and so fidelity
susceptibility should have maximum there. One can use Eq. (\ref{chii}) to
easily analytically compute the location of the maximum of fidelity
susceptibility and quantify how fast it approaches $g_c=1$ as the size of the
system is increased. One can also study with the help of Eq. (\ref{chii}) 
the dependence of fidelity susceptibility 
on the system size and the distance from the critical point, which yields 
further insights into the quantum criticality of the Ising model (see Ref. \cite{BD_fidPRE} for
the details).

{\bf Counterdiabatic driving.} Perfectly adiabatic evolutions of a quantum
system subjected to arbitrary time variation of an external field 
can be engineered by a proper modification of a Hamiltonian 
\cite{DemirplakJPC2003,BerryJPA2009,AdolfoAdv2013,AdolfoPRX2014}.

For the quantum Ising chain this can be obtained after adding to the Hamiltonian (\ref{HIsing}) 
\bee
\Delta\hat H=-\dfrac{dg}{dt}\B{\sum_{m=1}^{N/2-1} h_m(g) \hat H^{[m]} + \frac{1}{2}
h_{N/2}(g)\hat H^{[N/2]}},
\eee
for even $N$ and 
\bee
\Delta\hat H=-\dfrac{dg}{dt}\sum_{m=1}^{N/2-1/2} h_m(g) \hat H^{[m]},
\eee
for odd $N$, where
\bxx
\hat H^{[m]}&=\hat H_{xy}^{[m]}+\hat H_{yx}^{[m]}, \  \
\hat H_{\alpha\beta}^{[m]}&=\sum_{n=1}^N \sigma_n^\alpha\B{\prod_{j=n+1}^{n+m-1}\sigma_{j}^z}\sigma_{n+m}^\beta
\exx
describes interactions of $m+1$ adjacent spins. This so-called counterdiabatic
Hamiltonian was computed for even-sized chains evolving in the positive parity subspace
in Ref. \cite{AdolfoPRL2012}. We have
checked that it can be written in such a form for systems of an arbitrary  size
evolving in a subspace of arbitrary parity.  

It should be perhaps explained that the fairly complicated counterdiabatic Hamiltonian
is obtained by ``reverse-engineering''. It is written down first in the
momentum space, where it takes a  simple form, and then it is 
transformed to the real space form.
It is thus assumed that the
system evolves in the subspace of the Hilbert  space of definite parity.
Such an assumption can be made because parity is conserved during  time evolutions with 
the counterdiabatic Hamiltonian: $[\hat H+\Delta\hat H,\hat P]=0$.

The strength of the ($m+1$)-body interactions in the counterdiabatic
Hamiltonian is given by 
\be
h_m(g)=\dfrac{1}{2N}\sum_{0<k<\pi} \dfrac{\sin(k)\sin(mk)}{g^2-2g\cos(k)+1}.
\label{zxcvb}
\ee
Using  mapping (\ref{g}) as well as  Eqs. (\ref{KA})--(\ref{KD}) and 
(\ref{B1})--(\ref{B4}),
one immediately finds that for any system size $N>1$ and $m=1,2,\cdots,N-1$ 
\bee
h_m(g)=\dfrac{g^{2m}\pm g^N}{8g^{m+1}\B{1\pm g^N}},
\eee
where the $+$ ($-$) sign should be used for evolutions taking place in the
positive (negative) parity subspaces of the Hilbert space. 
The closed-form expression for Eq. (\ref{zxcvb})  in even-sized systems evolving in the positive
parity subspace was previously reported in Ref. \cite{BD_counter}.

Summarizing, we have derived several  expressions for the sums leading to 
hyperbolic functions. They 
generalize and extend the results listed in the popular 
Gradshteyn and Ryzhik book \cite{Ryzhik}. We expect that they will be useful
in different studies and illustrated their applicability to the quantum Ising
model, which has provided the primary motivation for this work.

This work has been supported by the Polish National Science Centre (NCN) grant DEC-2013/09/B/ST3/00239.
I would like to thank Marek Rams for his comments about this article.


\end{document}